# Anomalous Power Law Decay in Solvation Dynamics of DNA: A Mode Coupling Theory Analysis of Ion Contribution

Biman Bagchi

**Solid State and Structural Chemistry Unit, Indian Institute of Science, Bangalore 560012, India.**

*Abstract*

Several time domain fluorescence Stokes shift (TDFSS) experiments have reported a slow power law decay in the hydration dynamics of a DNA molecule. Such a power law has neither been observed in computer simulations nor in some other TDFSS experiments. Here we observe that a slow decay may originate from collective ion contribution because in experiments DNA is immersed in a buffer solution, and also from groove bound water and lastly from DNA dynamics itself. In this work we first express the solvation time correlation function in terms of dynamic structure factors of the solution. We use mode coupling theory to calculate analytically the time dependence of collective ionic contribution. A power law decay in seen to originate from an interplay between long range probe-ion direct correlation function and ion-ion dynamic structure factor. Although the power law decay is reminiscent of Debye-Falkenhagen effect yet solvation dynamics is dominated by ion atmosphere relaxation times at longer length scales (small wave number) than in electrolyte friction. We further discuss why this power law may not originate from water motions which have been computed by molecular dynamics simulations. Lastly, we propose several experiments to check the prediction of the present theoretical work.

**Keywords:** DNA hydration dynamics, Power law decay, Debye-Falkenhagen effect, wave number dependent ion atmosphere relaxation

# I. Introduction

Significant progress has been made in understanding of molecular relaxation processes in the liquid phase in the last three decades [1]. Unfortunately, however, our progress towards developing a quantitatively accurate theory of electrolyte solutions has been less than satisfactory. This is because of the long range nature of ion-ion and ion-water interactions that introduces several severe complexities which hindered progress. As a result, our ability to calculate such things as the transport properties of aqueous electrolyte solutions has remained somewhat imperfect even after a century of study, although notable progress has been made in recent years by efforts of many scientists, notably by Turq and co-workers. [2-7] One of the major difficulty of understanding has been the unavailability of experimental probes that could explore electrolyte dynamics at different length scales. The transport properties that we usually address are concentration dependence of the bulk, zero frequency properties, such as conductivity and viscosity. The latter, although not hard to measure experimentally, is quite hard to calculate theoretically. More recently, attention has been focussed on time dependent properties at short time scales that gave rise to interesting information [8-13]. The short time properties of the electrolyte are however strongly coupled with dynamics of these solvent molecules. As water is most often the solvent, complex dynamics of water make study of properties of electrolyte solution quite a formidable problem. Although several computational studies of long time dynamics have appeared, these are limited by the following two factors. (i) There is hardly any atomistic simulation study that explores dynamics beyond the time scale of 1 ns. This is an important time range because the time for atmospheric relaxation, $\tau_{atm}$, is also of the order of ns in dilute electrolyte solutions. Computer simulation of dilute electrolyte solutions is hampered again by two factors. (a) A large system size is required to simulate dilute electrolyte solutions and (b) long time simulations are required to capture relaxation beyond 1 ns. Actually one may need simulations beyond many tens of ns. (ii) One would need to study several concentrations and different electrolyte solutions as different electrolytes may show quite different properties. For example, the well-known anomalous concentration dependence of viscosity of KCl solution is yet to be explained quantitatively. NaCl solution shows no such anomaly in the concentration dependence of viscosity. Such effects cannot be explained by using electrohydrodynamic description often employed to describe transport processes in electrolyte solutions. [14-18]

On the theoretical study, a fully dynamically self-consistent treatment of electrolyte transport properties at low to intermediate concentration is not yet fully developed. Turq and co-workers made pioneering advances employing the mean spherical approximation (MSA) and their treatment provide description of the concentration dependence of electrolyte conductivity of concentration till about 1M. [2-7] Despite many successes, this treatment was not developed to treat dynamical properties like coherent dynamical structure factor and visco-elastic properties like frequency dependent viscosity. More recently, such an approach was initiated by using a mode coupling theory (MCT). [9-13, 19] In MCT, ion-ion dynamical correlation terms are evaluated by using two slow variables which are the charge density and the current density terms. MCT leads to self-consistent expressions for frequency dependent friction on a tagged ion (discussed below). The friction is naturally time dependent because of two different contributions (ion atmosphere and electrophoretic) that relax with different time constants. [10] The most important advantage of MCT is that it provides a set of self-consistent equations that can be evaluated quantitatively. [10] The mode coupling theory approach provides a simple and microscopic derivation of (a) Debye-Huckel-Onsager law of concentration dependence of ionic conductivity, [10, 12] (b) Debye-Falkehagen law of frequency dependence of electrolyte conductance [11] and Onsager-Fuoss expression of the concentration dependence of viscosity. [12] In addition, it provides an explanation of the different experimental observations on ion diffusion constant reported by QENS and NMR measurements. [13]

Many of the detailed predictions of the MCT however have gone untested till now. For example, the prediction of a slow decay of intermediate scattering function [20, 21] at times comparable to atmosphere relaxation time ($\tau_{atm}$) has not been tested. As already mentioned, this is difficult to test by simulation. Experimental observation has been scarce except the indirect evidence of time dependent diffusion study reported in Ref.13.

An interesting way to study electrolyte dynamics would be put a probe inside an electrolyte solution and study the time dependent response of the probe to a sudden change in the charge distribution or polar properties of the probe, executed by optical excitation. While short time response of the medium will involve contributions from both solvent and ions, the long time contributions (those measuring beyond 100ps) are expected to be dominated by such experiments were carried out by several groups [22-28], with mixed success. One important limitation of some of the earlier studies was that the measurements were limited to relatively shorter time scales. Here the difficulty lies in the fact that solvation in bulk water is essentially complete within a few ps, so even a time scale of 10 ps appeared to be longer. [29, 30] However, the time scales involved in electrolyte solutions can be much longer as we discuss below.

Motion of a DNA (or, motion of a protein along a DNA) is coupled to motions of ions and water molecules. In order to understand this coupling, ultrafast laser spectroscopy has been used with time resolution extending from a few fs to tens of ns. The often employed technique is time dependent fluorescence Stokes shift (TDFSS) where a suitably placed dye molecule is excited optically and the shift in the frequency of the emission spectrum is measured as a function of time. In the case of DNA solvation dynamics, two different approaches have been adopted. [22-29] In one approach, a base of the DNA duplex is replaced biochemically by a dye molecule, such as coumarin or aminopurine. [29] In another approach a fluorescent probe is intercalated to one of the grooves of the DNA. [22-28] While these two approaches do not give identical results, some common features have been detected. Prominent among them is the existence of an apparently universal power law decay component, first reported by Berg and co-workers. [22-25] However, when probed with ultrafast time resolution (of the order of ps) or in computer simulation studies with run time of the order of 1 ns or so, the power law decay is not observable. It seems to appear in the longer time. [29, 30]

The power law seems to exist over four to five decades of time, typically arises after a substantial initial decay but appears when the amplitude of correlation is still significant. The origin of this power law decay component has remained ill-understood, although it has been attributed tentatively to the relaxation of ions in and around DNA. [22-25]

As already mentioned the experiments of Zewail, Pal and co-workers did not find evidence of any slow decay where they used amino purine as a substitute base to probe the solvation dynamics around DNA. [29] Computer simulations [30] are in good agreements with the results of zewail and co-workers. More recently, Corcelli and co-workers pointed out that a slow decay of the timescale of ns can indeed originate from a large amplitude conformational fluctuation of a damaged DNA. [28] In this case a transition of closed and opened form of DNA can produce the observed timescales. There are two issues that remained unclear or unsettled: The studies of Corcelli et al. (which consist of experiments and simulations) do not easily explain the origin of the observed power law because the transition between two conformations is expected to introduce a slow quasi-exponential decay unless it is coupled to solvent (water molecules) and ions to involve a range of timescales. Second, it is not clear that the experiments of Berg, Sen and co-workers employ a DNA that is seriously damaged, particularly when the probe is located in the groove. [22-27] It seems fare to assume that DNA is not damaged in the studies performed by Zewail and co-workers. [29]

Thus the issue of the power law remains largely unresolved.

Power law decays are of course well-known in dynamical critical phenomena, and also near weakly first order phase transition, as in isotropic-nematic phase transitions [31, 32]. In such cases, the power law decay arises from a softening of free energy surface near a phase transition. Power law decay is also routinely observed in supercooled liquids near glass transition where the origin is attributed to correlated dynamics. Power law decays can be considered as a limit of Kohlrausch-William-Watts (KWW) stretched exponential form with the value of the stretching exponent (usual notation β) becomes small. [33, 34] While a stretched exponential form of decay can sometimes be explained in terms of a distribution of relaxation times, a power law decay is hard to explain in the same fashion. Usual explanation of power law near a critical point and in supercooled liquid employs a mode coupling theory argument where slowing down of density relaxation ( could be due to softening of free energy surface for density fluctuation) coupled to other hydrodynamics (or, sometimes non-hydrodynamic) modes can be to give rise to the observed power law decay. [20, 21]

In DNA solution, there could be several alternative sources of slow decay, although it is by no means clear that any of them can give rise to power law decay. The first candidate of course is the water that is bound to grooves of DNA, in particular those bound to the minor groove of water. In the AT-rich minor grooves these water molecules form a "spine of hydration" that is ice-like with enhanced tetrahedral ordering [35]. These water molecules can certainly make slow contribution, although need not be like a power-law.

The second source of power law decay is the ions. In a DNA solution, there is always a buffer solution which is often sodium phosphate. The solution is in the 100-500 mM range which is pretty significant due to the long range nature of ion-dipole interaction. [36] This contribution is in addition to the contribution of the counter ions. This contribution could involve distortion of the heterogeneous ion distribution that exists around a DNA, as mentioned above. Another issue is that the phosphate ions are pretty large, giving rise to low self-diffusion coefficient. Thus, the ion atmosphere relaxation time, given by $\tau_{atm} = \frac{1}{D_I \kappa_D^2}$, where $D_I$ is self-diffusion of an ion and $\kappa_D$ is inverse Debye screening length, has a value between 1-10 ns in the said buffer concentration range (0.01M-0.05M solution).

A third possibility is the contribution of DNA chain itself. But in the experimental studies mentioned above the chain is too short and as a result the DNA chain should be still rigid. Thus, this contribution seems to be less likely in the present context.

In this paper we propose a microscopic explanation of the origin of this power law by using a molecular theory of solvation dynamics that includes contribution of the ion atmosphere relaxation of DNA solution. The molecular theory that we employ is based on time dependent density functional theory of statistical mechanics and can be cast in the form of a mode coupling theory expression and also has been successfully used in several occasions to understand observed non-exponential salvation dynamics in dipolar liquids and also in ionic liquids.

In essence, the power law owes its origin to essentially the same mechanism that is responsible for Debye-Falkenhagen effect well-known in classical electrochemistry. We are aware that theoretically this effect has been criticised as giving the wrong time scale. Two comments are in order. First, what determines solvation dynamics is wave number dependent dynamics and 2nd is frequency dependent.

When a probe attached to DNA is optically excited, a new charge distribution is created in the probe solute. Subsequent stabilization of this new charge distribution may involve contribution from DNA base pairs, negatively charged phosphate groups, counterions and also the negatively charged ions present, which in the most cases are phosphate and biphosphate ions.

In the reported experiments, one uses a very small piece of DNA, consisting of 12 to 16 base pairs. In such small systems, the negatively charged phosphate ions and the bases (which have dipole and quadrupole monets) are rather rigidly held. They do not have any long wavelength or large amplitude motions to contribute significantly to slow decay observed in experiments and are unlikely players in the same. This leaves water, counterions and negatively charged ions as the probable cause for the slow decay. Water molecules can rotate and translate to solvate the charged probe solute. Ion atmosphere relaxation in the electrolyte solution can also couple to salvation dynamics. While this aspect has obviously been realized, no quantitative account of this effect has been taken into account. [19, 22-25]

To summarize our discussion so far, we have pointed out that while in simulations of DNA solvation dynamics, the system contains only the negatively charged DNA, positive counter ions (mostly Na+) and water molecules; in real experiments a DNA solution contains both positively and negatively charged ions, as a buffer is used to stabilize the DNA. The buffer typically used is a solution of sodium phosphate solution consisting of $Na_3PO_4$ and $Na_2HPO_4$. These ions can give rise to slow relaxation in th ns time domain and at the same time can explain the power law decay, as we discuss below.

**II. Theoretical Formulation**

In most of the solvation dynamics experiments, one measures the time dependent Stokes shift of the emission spectrum. Solvation time correlation function is so normalized that it is unity at time t=0 and goes to zero as time increases to infinity. [37] Time dependent energy of the solute probe energy is usually denoted by average frequency ν(t) of the emission spectrum

$$C_S(t) = \frac{\nu(t)-\nu(\infty)}{\nu(0)-\nu(\infty)}, \tag{1}$$

In the theoretical studies one invokes linear response theory to equate the non-equilibrium time correlation function measured in experiments to energy-energy time correlation function evaluated at equilibrium and defined by [37]

$$C_{EE}(t) = \frac{<\delta E(0)\delta E(t)>}{<\delta E(0)\delta E(0))>}, \tag{2}$$

Under linear response theory, $C_S(t) = C_{EE}(t)$. We shall henceforth refer to salvation time correlation function as $C_S(t)$. Theoretically it is relatively simpler to calculate $C_{EE}(t)$.

In the case of DNA salvation dynamics, the experimental results can then be expressed in the following form

$$C_S(t) = \sum_{i=1} A_i \exp(-t/\tau_i) + (1-\sum_i A_i)t^{-\alpha}, \tag{3}$$

This expression describes the fact that there is a crossover from exponential-like decay to a power law decay of the salvation time correlation function. Strictly speaking, there is also a Gaussian time dependent term at short times, but we shall ignore it in the present discussion as we are interested in the long time dynamics of salvation.

### A. Expression for solvation energy

In order to understand the role of these ions we start with the following expression of salvation energy of a probe located at position **r** and at time t, provided by the time dependent density functional theory of statistical mechanics

$$E_{p,solv}(\mathbf{r},t) = -\int d\mathbf{r}' \sum_i c_{p,i}(\mathbf{r}') \, \delta\rho_i(\mathbf{r}-\mathbf{r}',t) - \int d\mathbf{r} \, c_{p,\omega}(\mathbf{r}',\omega) \, \delta\rho_w(\mathbf{r}-\mathbf{r}',\omega,t) ,$$

(4)

where $c_{p,i}$ is the probe - ion direct correlation function and $\rho_i(r,t)$ is the density of the ion i. The sum is over +ve and –ve ions We can include the contributions of other base pairs in a similar fashion, but neglected here, as normal modes of decay that is used to describe DNA dynamics should decay on a much faster time scale. [38] As already mentioned, the above expression has been derived by using time dependent density functional theory, and such expressions can be reduced to continuum model expressions.

Eq.3 can be used to obtain the following expression for the energy-energy time correlation function

$$<\delta E(0)\delta E(t)> = \int dk\, k^2 \sum_{i,j=1}^{2} c_{pi}(k)c_{pj}(k)F_{ij}(k,t) + \int dk\, k^2 [c_{pw}^2(110,k)F_w(110,k,t)] \\ + \int dk\, k^2 [c_{pw}^2(111,k)F_w(111,k,t)]$$

(5)

where $c_{pi}(k)$ is the Fourier transform of i-th ion-probe wave number dependent direct correlation function, $c_{pw}(11m;k)$ are the probe-water direct correlation function, expanded into spherical harmonics, $F_{ij}(k,t)$ are the partial intermediate scattering function between ion of type i and of type j. $F(11m;k,t)$ are the spherical harmonic expansion of the angle dependent intermediate scattering function of liquid water.

The above rather complex expression simplifies considerably in the long time because the last two terms in the above expression decays on a much faster time scale as they involve rotational motion of water molecules. Water molecules even in the grooves of DNA rotate rather fast, on the time scale of tens of ps. These motions are of course important to understand salvation dynamics in the ultrafast (sub-ps) to intermediate time scales (of the order of tens of ps) but not expected to make any significant contribution to the slow times (of the order of ns) where the power decay is observed. We therefore simplify the above expression by keeping only the first term that involves slow motion of ions.

Next, we assume that the probe can be approximated by a point dipole. Under this approximation, $c_{p+}(k)$ and $c_{p-}(k)$ are the same. We can then combine the two terms

$$<\delta E(0)\delta E(t)> = A[\int_0^\infty dk\, k^2 c_{p+}^2(k) F_{++}(k,t) + \int_0^\infty dk\, k^2 c_{p-}^2(k) F_{--}(k,t) + 2\int_0^\infty dk\, k^2 c_{p+}(k) c_{p-}(k) F_{+-}(k,t)]$$

(6)

where A is a numerical constant. We now proceed to obtain an asymptotic expression for the above integral by evaluating the long wave length and long time limit of the above integral. While the ion-dipole direct correlation function is long ranged, it is the evaluation of the intermediate scattering function $F_{++}(k,t)$ of the ions that requires special treatment and is non-trivial to evaluate. Here we combine time dependent density functional theory and mode coupling theory to obtain this function.

### B. Dynamic structure factor

Dynamics of electrolyte solutions are often described in terms of theories initiated by Debye, Huckel and Onsager (with notable contributions by Fuoss and Falkenhagen). Later statistical mechanical theories have been used to understand the dynamics. The classical theories can be used to obtain the equilibrium partial structure factors, $S_{++}(k)$, $S_{+-}(k)$ and $S_{--}(k)$. However, evaluation of the dynamic structure factors or the intermediate scattering functions requires special treatment.

In the time dependent density functional theory approach to electrolyte dynamics, one uses a molecular hydrodynamic approach where two additional conserved quantities, charge density and charge current are included in addition to the usual density and the current density.

Such an exercise was carried out in a series of papers [8, 10-13]. The theory so developed lead to a microscopic derivation of Debye-Huckel-Onsager expression of electrolyte conductivity, Onsager-Fuoss expression of concentration dependence of viscosity and also Debye-Falkenhagen expression. In the derivation of three celebrated expressions, use was made of established statistical mechanical expression of the transport coefficients, put them into the mode coupling theory forms and then evaluated by using an expression for the dynamic structure factor.

Thus, the central to the theory is the dynamic structure factor, as discussed above. As evident from our discussion of solvation dynamics, ion-ion dynamic structure factor also determines the long time behaviour of solvation time correlation function.

We now proceed to present an expression of the structure factor. The main aspects have been discussed in Refs. (9-11), so we shall present here the essential details.

General considerations based on time dependent statistical mechanics lead to the following expression for the dynamic structure factor $S_{++}(k,\omega)$ in the overdamped limit, for symmetric monovalent ions,

$$S_{++}(k,z) = \frac{S_{++}(k)}{z + \frac{D_+(z)k^2}{S_{++}(k)}}, \qquad (7)$$

where $z=-i\omega$, $D_+(z)$ is the frequency dependent self-diffusion coefficients of the cation, and $S_{++}(k)$ is the partial static structure factor among cations. Eq.7 is an approximation. This expression is valid only under the approximation that the anions and cations are identical in every respect, and that the solvent is a structure less continuum. That is, we assume the primitive model of electrolyte solution. Such an assumption is valid in the overdamped and long time limits.

We can use mode coupling theory to obtain an expression for $D_{++}(z)$ in the following fashion. First we write generalized Einstein's relation between diffusion coefficient and friction

$$D_+(z) = \frac{k_B T}{z + \zeta_+(z)}, \qquad (8)$$

Where $\varsigma_+(z)$ is the frequency dependent friction acting on the positive ion.

MCT is now used to calculate the ionic friction. First, it is decomposed into Stokes friction and electrolyte friction

$$\zeta_+(z) = \zeta_{Stokes} + \zeta_{elec}(z) , \qquad (9)$$

where we have neglected the frequency dependence of the Stokes friction due to viscosity. [19] We have invoked a time scale separation inherent in writing the above expression. The Stokes friction due to viscosity is expected to be frequency independent in the range below $z=10^{10}$ s$^{-1}$ or so. Atmospheric relaxation on the other hand is slow and $\tau_{atm}$ can be large, in the range of ns or longer. For example, for monovalent ions at 0.1M concentration, ion atmosphere relaxation is 1 ns. Thus, the above expression reflects the separation of time scales between ion atmosphere relaxation and local density (and also momentum which has even faster) relaxation.

As is well-known, the electrolyte friction consists of two contributions --- electrophoretic term and ion atmosphere term. MCT can be used to calculate both these terms, by using slow variables as charge density and current density, as was discussed in Refs. [9-11].

However, the electrphoretic term is significant only at high frequency (as it is connected with charge current density) where $z\tau_{atm}$ is much larger than unity. So, the electrophoretic can be neglected at times that are comparable to $\tau_{atm}$.

MCT provides the following expression for the friction due to the ion atmosphere term

$$\zeta_{elec}(t) = \frac{k_B T}{3(2\pi)^3} \int d\mathbf{k}\, k^2 [C_S(k)] \otimes [F(k,t)] \otimes [C_S(k)]^\dagger F_S(k,t) , \qquad (10)$$

In this compact notation, $[C_S(k)]$ is a row matrix defined by,

$$[C_s(k)] = \left[ \sqrt{\rho_1} C_{S1}(k) \sqrt{\rho_2} C_{S2}(k) \right] \quad , \tag{11}$$

Where the subscript "S" stands for ions that can be positive or negative, and $[C_S(k)]^\dagger$ is the transpose of $[C_S(k)]$. $[F(k,t)]$ is the 2X2 intermediate scattering function matrix with elements $[F_{\alpha\beta}(k,t)]$, $\alpha, \beta = 1, 2$. Clearly, $[F(k,t)]$ becomes the structure factor matrix $[S(k)]$ at $t = 0$.

There are several comments that are in order. First, expression for friction is similar to that for salvation energy. That is because one is energy-time correlation function, and the other is force-force correlation function. Both are derived from electrolyte density fluctuations Second, in this work positive and negative ions are assumed to be identical in all respects except charge. Thus, the direct correlation matrix $[C_S(k)]$ that simplifies considerably and the matrix elements become $\rho C_{S1}(k) C_{S2}(k)$. Then the expression of the direct correlation function, given below; they differ only by sign.

We assume that the ions are point ions and use Debye–Huckel (DH) theory of ion–ion pair correlations. The ion–ion partial structure factor is then given by

$$S_{\alpha\beta}(k) = \delta_{\alpha,\beta} - \frac{4\pi q_\alpha q_\beta \sqrt{\rho_\alpha \rho_\beta}}{\varepsilon k_B T} \frac{1}{k^2 + \kappa_D^2} \quad , \tag{12}$$

where the inverse Debye screening length $\kappa_D$ is defined by $\kappa_D^2 = \frac{4\pi \rho q^2}{\varepsilon k_B T}$. The ion–ion direct correlation function for the point ions in DH theory is given by

$$C_{\alpha\beta}(k) = -\frac{4\pi q_\alpha q_\beta}{\varepsilon k_B T} \frac{1}{k^2} \quad , \tag{13}$$

If we assume that the relaxation of the ionic van Hove functions is described by diffusional motion, that is,

we ignore the frequency dependence of $D_\alpha(z)$ in Eq.8 and replace it by its zero-frequency value $D_\alpha(\omega)$ and solve the resultant equation in the time domain, then we obtain a simple exponential decay for the dynamic structure factor, given by

$$[F(k,t)] = S(k)\exp\left(-[D]k^2 t [S(k)]^{-1}\right), \qquad (14)$$

where $[D]$ is the diagonal matrix of self-diffusion coefficients. The above expression is certainly an approximate one, but it describes a substantial part of the decay of $F(k,t)$ (may be up to 80% or so) and at the same time simple enough to use in analytical approach. When Eqs. (11-14) are substituted in Eq. 10, the resultant integral over the wave vector **k** can be evaluated analytically [11] and the final result of the time-dependent microscopic electrolyte friction is given by

$$\zeta_{elec}(t) = \frac{q_s^2 \kappa_D^2}{3\varepsilon}\left[\frac{e^{-D\kappa_D^2 t}}{\sqrt{2\pi D t}} + \kappa_D e^{D\kappa_D^2 t}\left\{\Phi\left(\sqrt{2D\kappa_D^2 t}\right) - 1\right\}\right], \qquad (15)$$

where $\Phi(x)$ is the error function and it is assumed that all ions have the same diffusion coefficient $D$. Note that $\zeta_{elec}(t)$ given by the above equation exhibits a non-exponential decay. It is also important to note that the above equation needs to be solve self-consistently as the diffusion coefficient on the right hand side depends on the total friction on the ions that include also the microscopic $\delta\zeta_{s,mic}(t)$ on the left hand side.

The Laplace transform of the above equation can be carried out analytically to obtain the following expression of the frequency-dependent friction:

$$\zeta_{elec}(z) = \frac{q_s^2 \kappa_D}{6\varepsilon D} \frac{1}{1 + \frac{1}{\sqrt{2}}\left[1 + z/D\kappa^2\right]^{1/2}}, \qquad (16)$$

One can thus express frequency dependent microscopic friction in terms of ion atmospheric relaxation time,

$$\zeta_{elec}(z) = \zeta_{elec}(0) \frac{1 + \sqrt{q}}{1 + \sqrt{q}\left[1 + z\tau_{atm}\right]^{1/2}}, \qquad (17)$$

Where q=1/2. $\tau_{atm}$ is the atmospheric relaxation time given by

$$\tau_{atm} = 1/D\kappa_D^2 \text{ where } \kappa_D^2 = \frac{4\pi\rho q^2}{\varepsilon k_B T}.$$

Here we have simplified the expression by setting $D_+ = D_-$ and $q = 1/2$ for symmetric identical ions. Here $\zeta_{elec}(0)$ is the zero frequency friction.

We now have, via generalized Einstein's relation, the following expression for ionic diffusion constant

$$D_+(z) = \frac{k_B T}{z + \zeta_{Stokes} + \zeta_{elec}(0)\frac{1 + \sqrt{q}}{1 + \sqrt{q}\left[1 + z\tau_{atm}\right]^{1/2}}}, \qquad (18)$$

Several comments on the above expressions are in order. (i) The unusual frequency dependence in $D_+(z)$ is a consequence of the long range nature of ion-ion interaction, and is responsible for Debye-Falkenhagen effect. (ii) As mentioned earlier, and emphasized here, there is a separation of time scales inherent in the above expression. The Stokes friction is frequency independent in the range below $z=10^{10}$ s$^{-1}$ or so. Atmospheric relaxation on the other hand is slow and $\tau_{atm}$ can be large, in the range of ns or longer. Thus, the above expression reflects the separation of time scales between ion atmosphere relaxation and local density (and also momentum which has even faster) relaxation. (iii) Thus, time dependent diffusion should show a decrease when plotted against time from one plateau given by $D/\varsigma_{Stokes}$ to $D/\varsigma_{Total}$, where $\varsigma_{Total}$ is the total friction, sum of Stokes and electrolyte friction. The frequency dependence should show arise as low values of frequency z are approached from above.

When Eq.18 is used in Eq.7, we find an expression for ion-ion dynamic structure factor that should exhibit power law temporal decay so long electrolyte friction is comparable to Stokes friction. If the electrolyte friction is much smaller than Stokes friction, then the decay shall be exponential, as expected. In reality, the dynamic structure factor will contain a short time decay which shall decrease the relative weight of both Stokes friction and electrolyte friction, as the overdamped limit is supposed to set in at longer times.

We are interested in the long time limit, of the order of ns. In that limit, the expression for $D_+(z)$ further simplifies to

$$D_+(z) = \frac{k_B T}{\zeta_{Stokes} + \zeta_{elec}(0)\dfrac{1+\sqrt{q}}{1+\sqrt{q}\left[1+z\tau_{atm}\right]^{1/2}}} \quad , \tag{19}$$

We consider a range $z\tau_{atm} \gg 1$. In such a limit, the above expression further simplifies to

$$D_+(z) \approx \frac{k_B T}{\zeta_{Stokes} + \zeta_{elec}(0)/\left[z\tau_{atm}\right]^{1/2}} \quad , \tag{20}$$

$$= \frac{D_{Stokes}}{[z\tau_{atm}]^{1/2} + \zeta_{elec}(0)/\zeta_{Stokes}} [z\tau_{atm}]^{1/2} ,  \qquad (21)$$

We now denote $\tilde{z} = z\tau_{atm}$ and denote $\alpha = \zeta_{elec}(0)/\zeta_{Stokes}$. Then we can re-write Eq.21 as

$$D_+(\tilde{z}) = \frac{D_{Stokes}\sqrt{\tilde{z}}}{\sqrt{\tilde{z}} + \alpha} , \qquad (22)$$

Here $D_{Stokes}$ is the ion self-diffusion coefficient in the limit of zero ion concentration. We shall use this expression in Eq. 7 of dynamic structure factor to study the asymptotic behaviour.

### C. Power law time dependence of dynamic structure factor

When Eq.18 is substituted in Eq. 7 we obtain

$$S_{++}(k,z) = \frac{S_{++}(k)}{z + \dfrac{k_B T}{[z + \zeta_{Stokes} + \zeta_{elec}(0)\dfrac{1+\sqrt{q}}{1+\sqrt{q}[1+z\tau_{atm}]^{1/2}}]} \cdot \dfrac{k^2}{S_{++}(k)}} , \qquad (23)$$

The non-Markovian character of this expression plays an important role in giving rise to a power law and also makes the situation quite different from Debye-Falkenhagen effect. This is a fact missed by earlier studies [39]. Since ionic contribution to solvation dynamics is dominated by low wave number ($k \sim 0$) modes and since the rate of decay of dynamic structure factor varies with $k^2$, *this rate could be smaller than the rate of ion atmosphere relaxation rate*. This in turn makes the relevant frequency z probed in the solvation dynamics to be small too.

The above expression is a bit too complex to evaluate analytically.

The dynamic structure factor is now given by

$$S_{++}(k,\tilde{z})=\frac{S_{++}(k)(\sqrt{\tilde{z}}+\alpha)}{\tilde{z}(\sqrt{\tilde{z}}+\alpha)+\dfrac{D_{Stokes}k^2}{S_{++}(k)}\sqrt{\tilde{z}}} \quad , \tag{24}$$

This expression has interesting structure. One can easily verify by Laplace inversion that F(k,t) exhibits a power law decay when $\alpha=\zeta_{elec}(0)/\zeta_{Stokes}$ becomes larger than $\tilde{z}=z\tau_{atm}$. This can happen at times longer than $\tau_{atm}$ so that frequency z explored is smaller than $1/\tau_{atm}$. In that limit, Laplace inversion gives the following expression

$$F_{++}(k,t)=S_{++}(k)\,[e^{\gamma^2 t}\,erfc(\gamma\sqrt{t})] \tag{25}$$

Where $\gamma=D_{Stokes}k^2/\alpha\,S_{++}(k)$.

It is important to note that the power law seems to be strong when $\alpha=\zeta_{elec}(0)/\zeta_{Stokes}$ is non-negligible. Clearly, the power law becomes important when the contribution of the electrolyte friction is comparable to (or, larger than) the Stokes friction.

However, even when the closed form expression (25) is not valid, Eqs.23 and 24 both predict power law decay in the time dependence of dynamic structure factor.

## III. Power law decay of solvation time correlation function

Eq.16 is rather informative. Let us look at certain limits to understand the general behaviour. In the k →0 limit, $S_{ij}(k)$ exhibits the following properties. $S_{++}(k)$ and $S_{--}(k)$ both goes to zero, strictly in the DH limit but otherwise also expected to become small. $S_{+-}(k)$ and $S_{-+}(k)$ on the other hand approaches unity. The large k behaviour is different in both the cases. Next, we consider the direct correlation function. $C_{p+}(k)$ and $C_{p-}(k)$ can have a divergent-like growth in the $k \rightarrow 0$ limit. If it retained $1/k^2$ behavior, then the solvation time correlation function could exhibit $(1/Dt)^{-1/2}$ type power law behaviour. However, the situation here is a bit more complex, and detailed calculations are required to find out the precise behaviour. This is because we need to consider in detail the k-dependence of $S_{+-}(k)$.

In the present case of probe stacked inside DNA or in minor groove, $C_{p+}(k)$ or $C_{p-}(k)$ are not expected to have infrared divergence but would certainly show a growth as we approach small k, but at the same time these functions themselves would go to zero in the $k \rightarrow 0$ limit. All these would combine to give rise to a dominant decay contribution from small k limit where relaxation is quite slow, slower than $1/D\kappa_D^2$ found for the electrolyte friction.

## IV.    Comparison with Experiments and Simulations

As mentioned earlier, there is an apparent lack of agreement between experimental and simulation studies about the existence of the power law. Computer simulations have failed to unearth it not only for solvation dynamics but also in pure electrolyte solutions, with the notable exception of the Brownian dynamics simulation of Dufreche et al. [13] which did find some signatures of power law decay but did not pursue it any further. [30] This absence is not difficult to understand. There can be a large separation of time scale between the short time and the long time contributions. The short time contribution is dominated by ion-water coupling and the decay time constant is less than 10 ps or so. [30, 22-25] If groove water is involved, even then the slow decay is not easily expected to (a) make a dominant contribution beyond 100 ps, (b) give rise to a power law with such a small power as observed by Berg and co-workers. [22-25] Thus, water does not provide an easy explanation of the power law decay.

Experiments of Zewail and Pal found decay with two time constants, [29] with the slowest one of the order of 20 ps, in agreement with the results of simulation, but in apparent disagreement with the experimental results of Berg, Sen and co-workers. [22-27] However, the former experiments did not explore the slow long time decay.

The apparently new observation made here is that all the experimental results employed a buffer solution of relatively high concentration. The buffer employed is usually sodium phosphate. These ions, in addition to the counter ions present naturally in a DNA solution, can make a noticeable contribution to the decay of the solvation time correlation function.

Our explanation, based on the theoretical work done here is as follows. In the DNA solution, there is an inhomogeneous distribution of ions created by the negatively charged phosphate ions of DNA. When we create a charge distribution in the solute probe, this inhomogeneous charge distribution needs to adjust by altering its distribution. This is a collective phenomenon involving collective response of the ions. As we have shown both the electrolyte friction and the coherent dynamic structure factor develop a slow power law decay behaviour in the long time, this redistribution of charges can also exhibit power law decay.

*The theory suggests the following experiments to check the explanation offered here*. First would be to study solvation dynamics of a probe like coumarin dye in the same buffer solution as employed in experiments with DNA solution. One can even use a simpler system of probe in an ordinary electrolyte solution like RbCl, but must look carefully at long times. Second, one should study solvation in DNA but with varying buffer concentration. If the long time decay is indeed due to collective dynamics of ions, then the amplitude of the power law should be affected by the concentration and the nature of the buffer. However, this dependence itself can be quite complex, as several important factors including atmospheric relaxation time, Debye length etc shall change with buffer concentration.

## V. Conclusion

Coupling of solvation dynamics with ion atmosphere relaxation has been discussed before, notably by van der Zwan and Hynes [39]. Such a study also predicts time scales comparable to $\tau_{atm}$. For 0.2 M concentration often used in experiments, this gives a time constant in the range of a few ns. The picture developed here is somewhat different. Here we show that $\tau_{atm}$ becomes further modified by interaction terms. Nevertheless, the origin of a slow term in the solvation dynamics was already addressed to in Ref.39.

The two dynamical features of electrolyte solution are the electrophoretic effect and the ion atmosphere relaxation effect. The first one occurs on faster time scale as it involves charge current relaxation. The latter, on the other hand, is a much slower process as it involves charge density relaxation. This is also related to Debye-Falkenhagen effect that describes frequency dependent conductivity of electrolyte solutions. It was shown elsewhere that Debye-Falkenhagen effect is indeed related to anomalous frequency dispersion of the electrolyte friction which in turn gives rise to anomalous frequency dependence of ion diffusion coefficient. [12] The power law decay obtained here is essentially the same as the Debye-Falkenhagen effect. However, the explanation offered here is different from Debye-Falkenhagen in that longer length scales are involved in solvation dynamics than in electrolyte friction.

As emphasized above, the existence of slow decay in an electrolyte solution is not surprising because the ion atmosphere relaxation effect is slow, much slower than other solvation processes (such as dipolar rotation) in a liquid. However, the existence of power law decay is non-trivial and has its origin essentially in the long range nature of ionic interaction.

In this work we show that solvation time correlation function, $C_S(t)$ for a solute probe in a binary electrolyte with oppositely charged ions of identical mass and size, in terms of an integral over the coherent dynamic structure factor (or, intermediate scattering function). We next show that the same electrolyte solution exhibits temporal power law decay in the time comparable to ion atmosphere relaxation time. This result has been used to explain the observation of power law decay in DNA solvation by exploiting the fact that a DNA solution uses a buffer which itself is an electrolyte.

Finally, we proposed several experimental investigations to check the predictions made here.

## Acknowledgement

It is a pleasure to dedicate this paper to Professor Pierre Turq on the occasion his 70th birthday. Pierre has been, for a long time, a valuable and helpful colleague and a trusted friend and great collaborator. We join his innumerable colleague in wishing him a Happy Birthday and a productive, healthy life. I thank Ms. Susmita Roy for discussions and for much help in preparing the manuscript. I thank Prof. S. Yashonath, Prof. S. Sen and Prof. S. K. Pal for helpful discussions. Particular thanks to Prof. Yashonath, De. Praveen Kumar and Ms. Susmita Roy for collaborations. We also thank an anonymous reviewer for many constructive criticism and suggestions, This work was supported in parts from grant by DST (India) and a Sir JC Bose Fellowship.

**REFERENCES**

1. B. Bagchi, *Molecular Relaxation in Liquids* (Oxford University Press, USA, 2012).

2. M. Moreau and P. Turq, *Chemical Reactivity in Liquids: Fundamental Aspects* (Plenum Press, New York, 1988).

3. P. Turq, J. Barthel, and M. Chemla, *Transport, relaxation and kinetic processes in electrolytes Lecture notes in Chemistry* (vol. **57**, Springer Verlag, Berlin, 1992).

4. S. Durand-vidal, J.-P. Simonin, and P. Turq, *Electrolytes at interfaces: Progress in Theoretical Chemistry and Physics* (vol. **1**, Kluwer Academic Publishers, 2002, 2nd ed.).

5. P. Turq, Chem. Phys. Lett. **15**, 579 (1972); A. Cadene, S. Durand-Vidal, P. Turq, and J. Brendle, Journal of Colloid and Interface Science, **285**, 719-730 (2005).

6. J. J. Molina, J. F. Dufrèche, M. Salanne, O. Bernard, M. Jardat and P. Turq, Phys. Rev. E **80** (6) 065103 (2009); J.-F. Dufrêche, O. Bernard, S. Durand-Vidal, and P. Turq, J. Phys. Chem. B, **109**, 9873-9884 (2005).

7. A. Carof, V. Marry, M. Salanne, J.P. Hansen, P. Turq et B. Rotenberg, Mol. Simul., **40**, 237 (2013); M. Salanne, C. Simon, P. Turq, and P. A. Madden, J. Fluorine Chem., **130** (1) 38-44 (2009); V. Dahirel, M. Jardat, J. F. Dufrêche, and P. Turq, J. Chem. Phys., **126**, 114108 (2007).

8. B. Bagchi, J. Chem. Phys. **95**, 467 (1991); B. Bagchi, J. Chem. Phys. **109**, 3989 (1998).

9. P. G. Wolynes, Ann. Rev. Phys. Chem. **31**, 345-376 (1980).

10. A. Chandra, R. Biswas, and B. Bagchi, J. Am. Chem. Soc. **121**, 4082 (1999).

11. A. Chandra and B. Bagchi, J. Chem. Phys. **112**, 1876 (2000); A. Chandra and B. Bagchi, J. Chem. Phys. **113**, 3236 (2000).

12. A. Chandra and B. Bagchi, Proc. Indian Acad. Sci. (Chem. Sci.) **101**, 83 (1989); A. Chandra and B. Bagchi, J. Phys. Chem. B **104**, 9067 (2000); A. Chandra and B. Bagchi, J. Chem. Phys. **110**, 10024 (1999).

13. J. F. Dufreche, O. Bernard, P. Turq, A. Mukherjee, and B. Bagchi, Phys. Rev. Lett. **88**, 95902 (2002).

14. S. Glasstone*, An Introduction to Electrochemistry Litton, New York,* (1942)*; H. S. Harned, The Physical Chemistry of Electrolyte Solutions Reinhold, New York,* (1958); *Ionic Interactions edited by S. Petrucci, Academic, New York*, (1971); H. Falkenhagen, Theorie der Elektrolyte Hirzel, Leipzig, (1971).

15. J. M. G. Barthel, H. Krienke, and W. Kunz, *Physical Chemistry of Electrolyte Solutions: Modern Aspects Steinkopff, New York*, (1998).

16. P. Debye and H. Huckel, Z. Phys. **25**, 49 (1924).

17. L. Onsager, Z. Phys. **27**, 388 (1926); **28**, 277 (1927).

18. L. Onsager and R. M. Fuoss, J. Phys. Chem. **36**, 2689 (1932)

19. B. Bagchi and S. Bhattacharyya, Adv. Chem. Phys. **116**, 67 (2001).

20. W. Götze, *Complex Dynamics of glass forming liquids. A mode-coupling theory. Oxford: Oxford University Press,* ISBN 978-0-19-923534-6 (2009).

21. W Gotze and L Sjogren, Rep. Prog. Phys. **55**, 241 (1992)

22. D, Andreatta, J. L. P. Lustres, S. A. Kovalenko, N. P. Ernsting, C. J. Murphy, R. S. Coleman, and M. A. Berg, J. Am. Chem. Soc. **127**, 7270 (2005).

23. E. B. Brauns, M. L. Madaras, R. S. Coleman, C. J. Murphy and M. A. Berg, Phys. Rev. Lett. **88**, 158101 (2002); E. B. Brauns, M. L. Madaras, R. S. Coleman, C. J. Murphy, and M. A. Berg, J. Am. Chem. Soc. **121**, 11644-11649 (1999).

24. D. Andreatta, S. Sen, J. L. Pérez Lustres, S. A. Kovalenko, N. P. Ernsting, C. J. Murphy, R. S. Coleman, and M. A. Berg, J. Am. Chem. Soc. **128**, 6885-92 (2006).

25. S. Sen, D. Andreatta, S. Y. Ponomarev, D. L. Beveridge, and M. A. Berg, J. Am. Chem. Soc. **131**, 1724 (2009); S. Sen, L. Gearheart, E. Rivers, H. Lui, R. S. Coleman, C. J. Murphy, and M. A. Berg, J. Phys. Chem. B **110**, 13248 -55 (2006); S. Sen, N. A. Paraggio, L. A. Gearheart, E. E. Connor, A. Issa, R. S. Coleman, D. M. Wilson III, M. D. Wyatt, and M. A. Berg, Biophys. J. **89**, 4129-38 (2005).

26. S. D. Verma, N. Pal, M. K. Singh, and S. Sen J. Phys. Chem. Lett 2012, **3**, 2621 (2012).

27. N. Pal, S. D. Verma and S. Sen, J. Am. Chem. Soc. **132**, 9277 (2010).

28. K. E. Furse and S. A. Corcelli, J. Am. Chem. Soc. **33**, 720–723 (2011); K. E. Furse and S. A. Corcelli, J. Phys. Chem. Lett. **1**, 1813 (2010); K. E. Furse and S. A. Corcelli, J. Am. Chem. Soc.**130**, 13103 (2008).

29. S. K. Pal, L. Zhao, T. Xia, and A. H. Zewail, Proceedings of the National Academy of Sciences, USA, **100**, 13746 (2003); S. K. Pal, L. Zhao and A. H. Zewail, Proceedings of the National Academy of Sciences, USA, **100**, 8113 (2003).

30. S. Pal, P. K. Maiti, B. Bagchi, and J. T. Hynes, J. Phys. Chem. B **110**, 26396 (2006); S. Pal, P. K. Maiti and B. Bagchi, J. Chem. Phys. **125**, 234903 (2006).

31. S. D. Gottke, D. D. Brace, H. Cang, B. Bagchi, and M. D. Fayer, J. Chem. Phys. **116**, 360 (2002); S. D. Gottke, H. Cang, B. Bagchi, and M. D. Fayer, J. Chem. Phys. **116**, 6339 (2002).

32. D. Chakrabarti and B. Bagchi, J. Phys. Chem. B (Feature Article), **111**(40), 11646 (2007).

33. R. Kohlrausch, Annalen der Physik und Chemie (Poggendorff) **91,** 56–82, 179–213 (1854).

34. G. Williams and D. C. Watts, Transactions of the Faraday Society **66**, 80–85 (1970).

35. B. Jana, S. Pal, and B. Bagchi, J. Phys. Chem. B, **114**, 3633 (2010); B. Jana, S. Pal, P. K. Maiti, S. Lin, J. T. Hynes, and B. Bagchi, J. Phys. Chem. B **110**, 19611 (2006).

36. S. Batabyal, T. Mondol, S. Choudhury, A. Mazumder and S. K. Pal, Biochimie **95**, 2168 (2013); S. K. Pal, personal communication.

37. B. Bagchi, Ann. Rev. Phys. Chem. 40, 115 (1989); S. Roy and B. Bagchi, Chem. Phys. **183**, 207 (1994); B. Bagchi and A. Chandra, Adv. Chem. Phys. **80**, 1 (1991); B. Bagchi and R. Biswas, Adv. Chem. Phys. **109**, 207 (1999); B. Bagchi and N. Gayathri, Adv. Chem. Phys. **107**, 1 (1999).

38. S. K. Pal, J. Peon, B. Bagchi, and A.H. Zewail, J. Phys. Chem. B **106**, 12376 (2002).

39. G. van der Zwan and J. T. Hynes, Chem. Phys. **152**, 169 (1991).